\documentclass[twocolumn,twoside,slac]{revtex4}
\usepackage{graphicx}
\usepackage{fancyhdr}
\def\babar{\mbox{\sl B\hspace{-0.4em} {\small\sl A}\hspace{-0.37em} \sl B\hspace{-0.4em} {\small\sl A\hspace{-0.02em}R}}}

\begin{document}

\title{Using Geant4 in the \babar\ Simulation}


\author{D.H. Wright, D. Aston, M.L. Kocian, H. Marsiske}
\affiliation{SLAC, Stanford, CA 94025, USA}
\author{W.S Lockman, D.C. Williams}
\affiliation{University of California at Santa Cruz, Institute for Particle Physics, Santa Cruz, CA 95064, USA}
\author{A.W. Weidemann, J.R. Wilson}
\affiliation{University of South Carolina, Columbia, SC 29208, USA}
\author{T.B. Moore}
\affiliation{University of Massachusetts, Amherst, MA 01003, USA}
\author{A.J. Lyon}
\affiliation{University of Manchester, Manchester M13 9PL, United Kingdom}

\author{on behalf of the \babar\ Computing Group}

\begin{abstract}
\babar\ was the first large experiment to incorporate Geant4 into its detector 
simulation.  Since July 2001, 1.5$\times$ 10$^9$ \babar\ events have been 
produced using this simulation.  In a typical 
$e^+ e^- \rightarrow \Upsilon (4s) \rightarrow B^0 {\overline {B^0}}$ event, 
between 30 and 60 tracks are produced in the generator and propagated through 
the detector, using decay, electromagnetic and hadronic processes provided by 
the Geant4 toolkit.  The material model of the detector is very detailed and 
a special particle transportation module was developed so that minute features
(on the few micron scale) would be sampled in the propagation without 
sacrificing performance.  The propagation phase for such an event requires 
5 CPU seconds on an 866 MHz processor.  Execution speeds for other \babar\ 
event types will also be presented.  Validation of simulated events with 
\babar\ data is ongoing, and results of Monte Carlo/data comparisons will be 
shown.  A discussion of the design of the simulation code, how the Geant4 
toolkit is used, and ongoing efforts to improve the agreement between data and
Monte Carlo will also be presented.  

\end{abstract}

\maketitle

\thispagestyle{fancy}

\section{OVERVIEW OF THE \babar\ SIMULATION}
\subsection{The Detector}
The primary physics goal of the \babar\ experiment is the study of CP violation
in the $B^0 \overline{B^0}$ system.  $B^0 \overline{B^0}$ events are produced
by the decay of $\Upsilon (4s)$ resonances which are produced in $e^+ e^-$
collisions.  Some of the $B^0$s decay into CP eigenstates such as 
$J/\Psi K^0_s$, and the subsequent decays of the $J/\Psi$ and $K^0_s$ are 
detected and reconstructed.  A typical event will produce between 30 and 60 
tracks, and typical energies of the final state decay products are:
\begin{itemize} 
\item lepton pairs: 0.3 $ < p < $ 2.3 GeV/c
\item $\pi^0$ : 0.3 $ < E < $ 2.5 GeV
\item $\gamma$ : 0.1 $ < E < $ 4.5 GeV .
\end{itemize}
Hadronic final states are also important because charged $\pi$s and $K$s
interact in the beam pipe and calorimeters.  Typical hadron momenta are
\begin{itemize} 
\item p $ < $ 4 GeV/c, with most $ < $ 1 GeV/c .
\end{itemize}
Background events such as Bhabha scattering produce the highest energy tracks,
with $ p_{e^-}, p_{e^+} <  9 $ GeV/c .

The \babar\ detector was built to collect final state tracks with high 
efficiency and precision, and to allow the reconstruction of $B^0$ decays
into a wide range of exclusive final states with low background.  The detector
was also designed to operate at the high luminosities provided by the PEP-II
accelerator.

The detector consists of a silicon vertex tracker (SVT) which surrounds the 
$e^+ e^-$ interaction point and provides vertex determination at the 10 $\mu$m
level.  Surrounding the SVT is a He-isobutane drift chamber (DCH) for the 
measurement of charged tracks.  Outside the DCH is a detector of 
internally-reflected Cherenkov radiation (DRC), constructed of quartz bars and
used for particle identification (PID).  Outside the DRC is a CsI 
electromagnetic calorimeter (EMC) designed to contain showers from charged and
neutral tracks.  A muon tracker and instrumented flux return (IFR), built of 
alternating layers of iron and resistive plate chamber (RPC) detectors, 
surrounds all the interior detectors.
  
\begin{figure*}[t]
\centering
\includegraphics[width=150mm, angle=0]{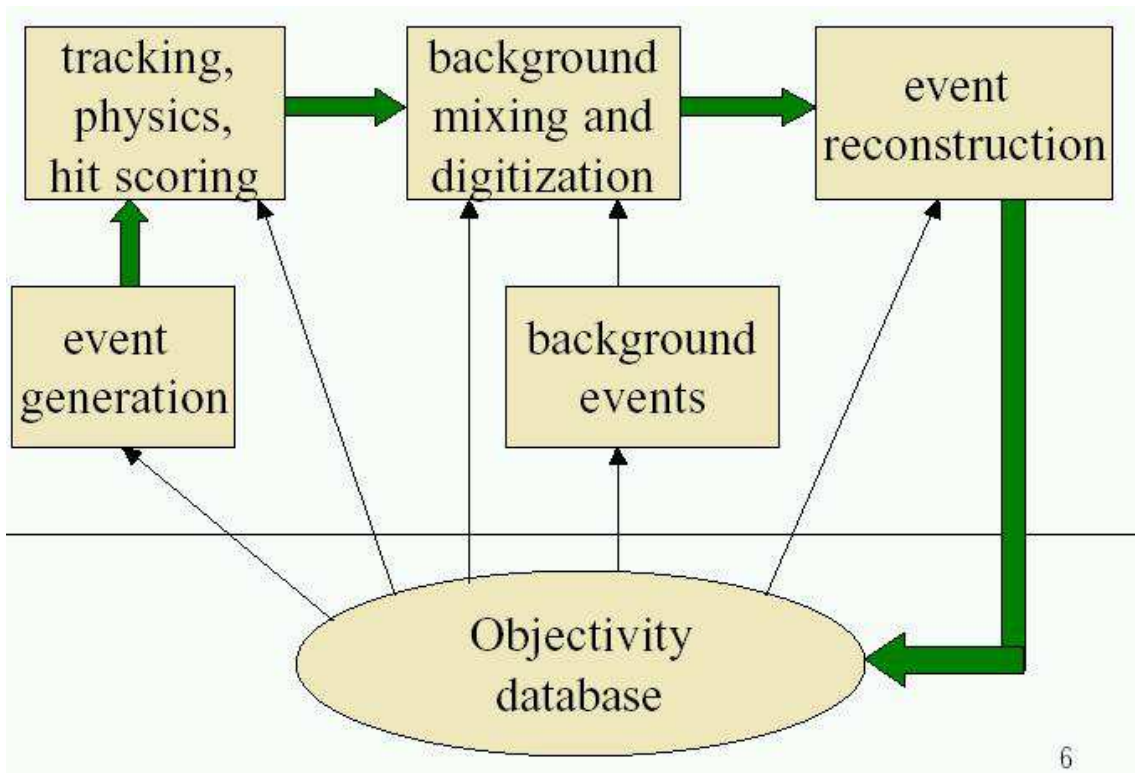}
\caption{Simple diagram of the \babar\ simulation.  Boxes above the horizontal 
line represent \babar\ Framework modules, and the wide arrows depict the 
event flow through the Framework.  Narrow arrows indicate the retrieval of 
information from the database.} \label{overview}
\end{figure*}

\subsection{The Simulation}
The simulation was designed to satisfy four main requirements.  It must:
\begin{itemize}
\item run within the \babar\ Framework \cite{Framework}.  All the tracking, 
physics and hit-scoring performed by Geant4 \cite{Geant4} is implemented as a 
Framework module.  Because the Framework is responsible for run and event 
control, Geant4 must therefore relinquish this duty.

\item work with existing event generators, detector response codes and 
reconstruction codes.  At the time Geant4 became available, \babar\ detector
response and reconstruction codes were essentially complete and many event
generation codes were already available.

\item use the Objectivity database \cite{Objectivity} for persistence.

\item be detailed, yet fast enough to keep up with high-luminosity event
production.
\end{itemize}

A simple diagram of the main functions and operation of the resulting 
simulation is shown in Fig.~\ref{overview}.

The Geant4 toolkit is used extensively in the simulation.  The detector 
geometry is built with Geant4 simple volumes and Boolean operations.  As 
mentioned above, Geant4 is also responsible for hit-scoring and provides
all the physics processes.  These include the decay, standard electromagnetic 
and low-energy hadronic ($E < 10 GeV$) processes.  Some features provided by
Geant4, such as detector response code and persistence, are not used.
The default Geant4 transportation/navigation process is also not used.  It 
employs a general Runge-Kutta stepper which was found to be too slow and not
precise enough for the many thin volumes in the \babar\ simulation.  Taking 
advantage of the slowly varying magnetic field in the \babar\ detector, a 
specialized stepper was developed to meet these needs.  This stepper 
determines how far a particle can travel before the field deviates from a 
locally constant value.  Perfect helices are then used to propagate the 
particle over that distance.  If that distance is large, many small steps are 
avoided.

\section{VALIDATION}

% \begin{figure*}[t]
\begin{figure*}[t]
\centering
\includegraphics[width=120mm, angle=0]{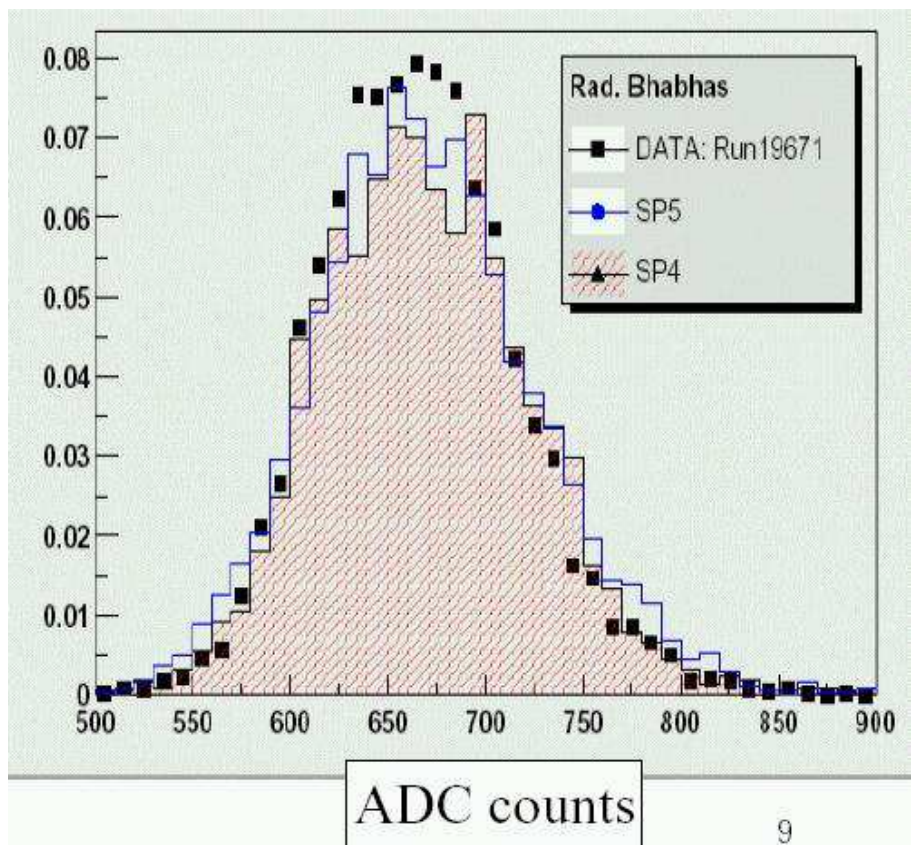}
\caption{$dE/dx$ for minimum ionizing electrons and positrons from Bhabha 
         scattering.  The histograms represent two different simulations of 
         the energy loss in the drift chamber gas.  The points are from a
         \babar\ data run.  The histograms are normalized to the number of 
         events in the data run.} \label{dedx}
\end{figure*}

Before being put into production, the \babar\/Geant4 simulation was subjected 
to a series of validation tests focusing on
 
\begin{itemize}
\item verifying the detector material model,
\item electromagnetic processes, 
\item hadronic processes,
\item tracking, resolution and reconstruction,
\item particle ID,
\item performance and robustness.
\end{itemize}

Since October 2000, 20 million simulated events of several types have been 
produced for these tests.  They include generic $B^0 \overline{B^0}$, Bhabha 
scattering and dimuon events, among others.  Comparsion of these events with  
data has allowed a refinement of the detector material model to the point 
that no further changes are necessary.  Understanding of particle
identification and electromagnetic processes is also believed to be well in 
hand, although validation in these areas continues.  However, additional 
validation is required for the hadronic processes.

% \begin{figure}[!ht]
\begin{figure*}
\begin{center}
\resizebox{6.2cm}{!}{\includegraphics{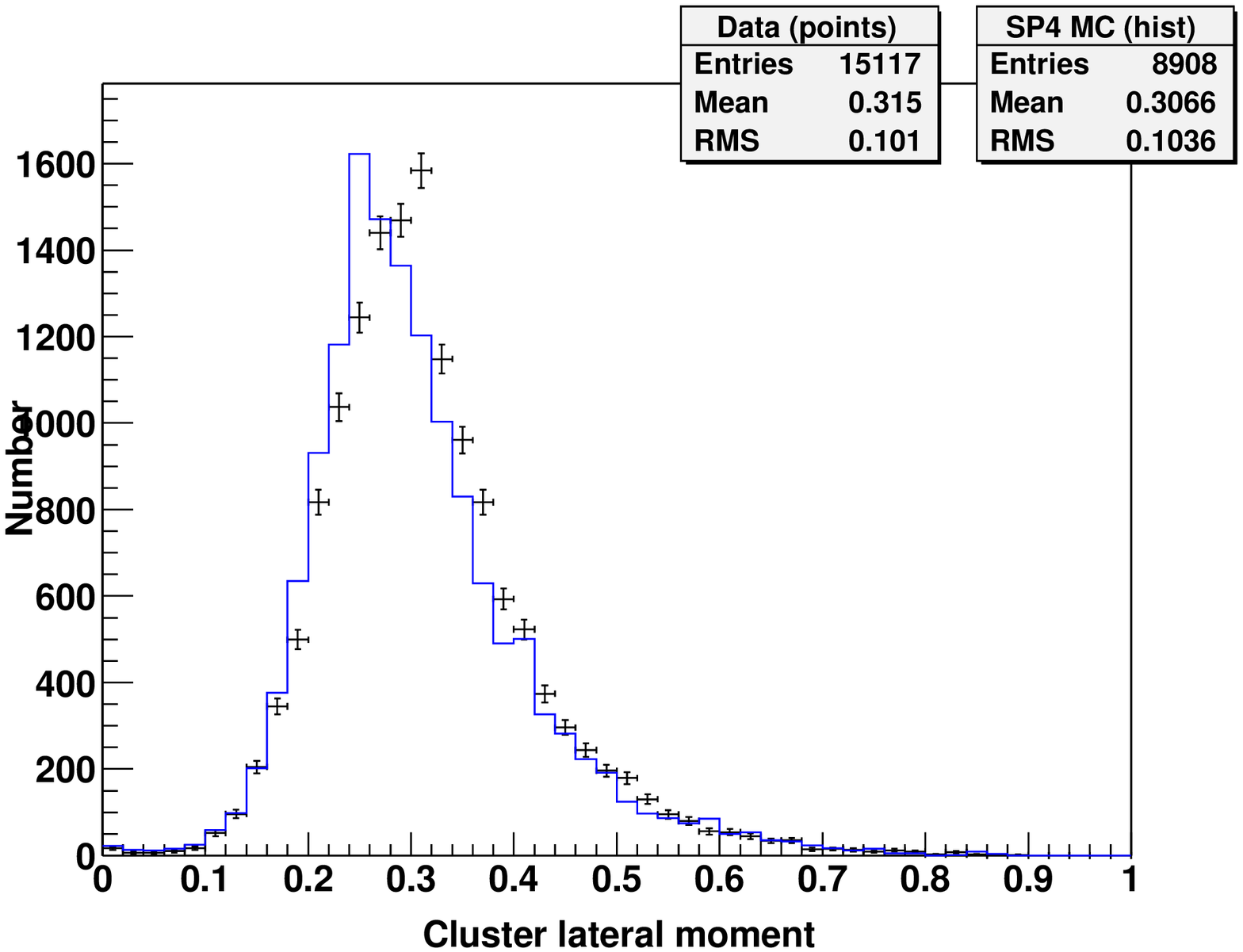}}
\resizebox{6.2cm}{!}{\includegraphics{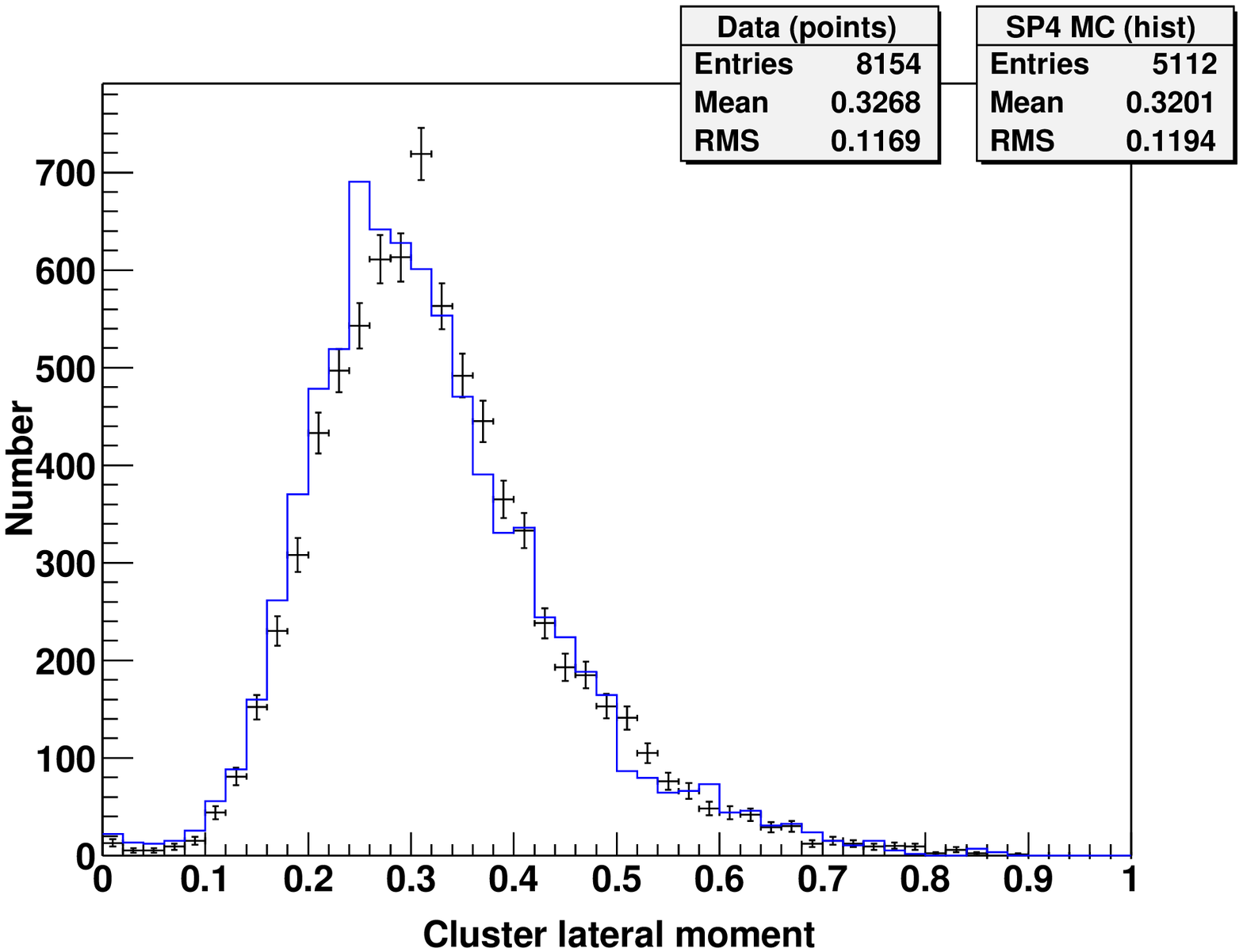}}
\resizebox{6.2cm}{!}{\includegraphics{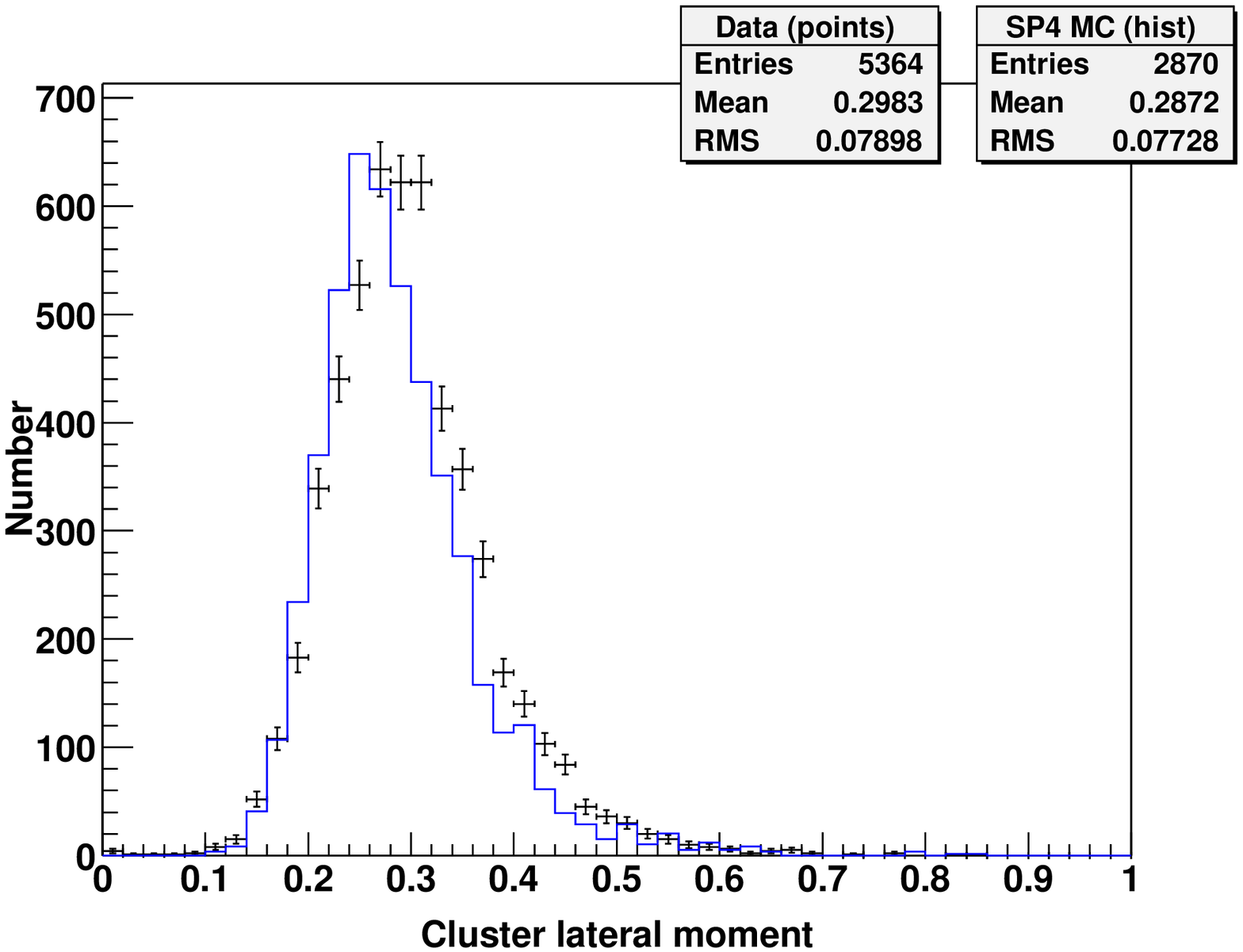}}
\resizebox{6.2cm}{!}{\includegraphics{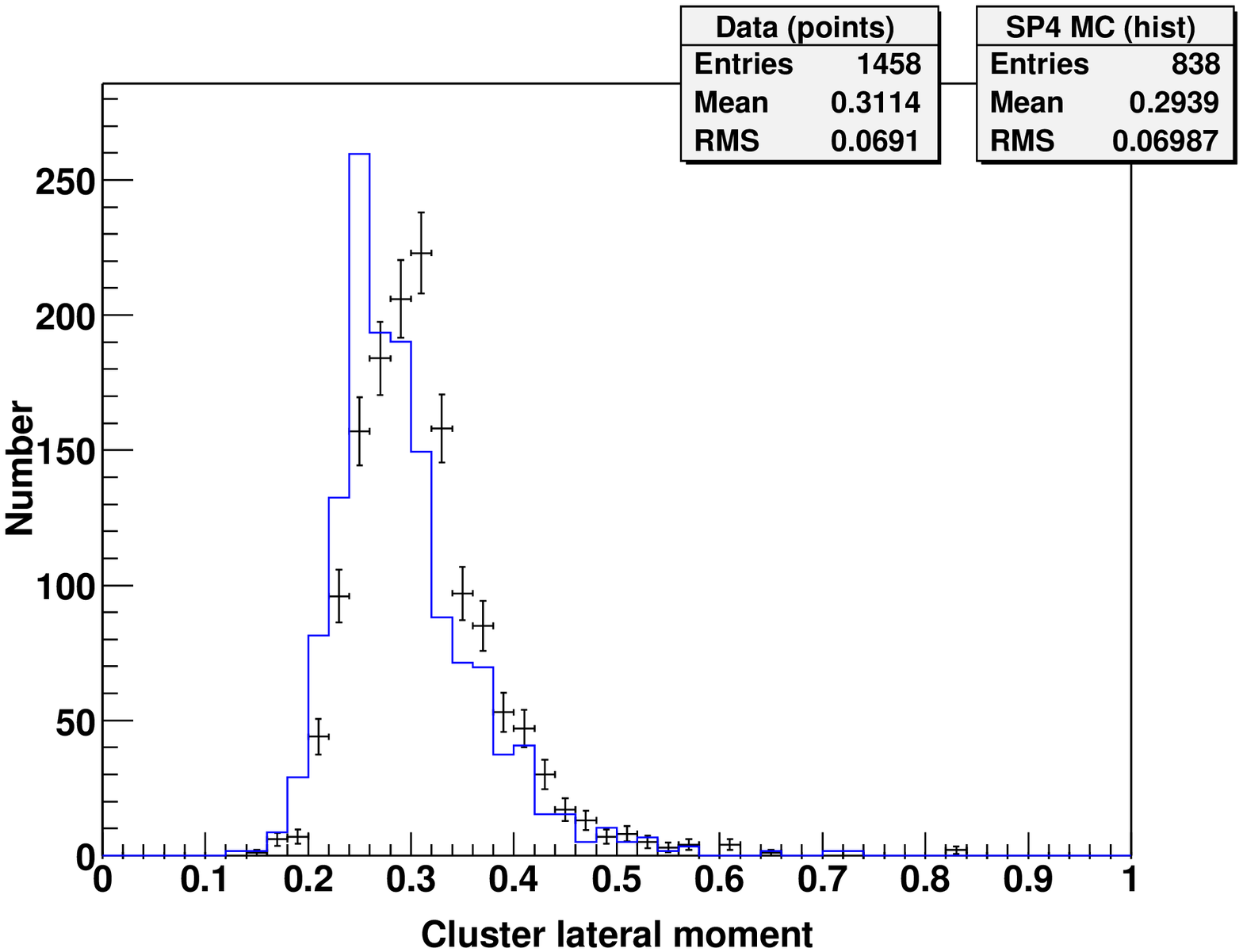}}
\end{center}
\caption{Distribution of the LAT parameter as a function of incident electron 
         or positron energy.  Top left: all energies;  top right: 0 to 1 GeV;
         bottom left: 1 to 3 GeV; bottom right: 3 to 5 GeV.  The histograms
         represent a simulated sample of Bhabha events and the points are 
         taken from \babar\ data.} \label{shower}
\end{figure*}
% \end{figure}

% \begin{figure*}
% \centering
% \includegraphics[width=120mm, angle=0]{shower.eps}
% \caption{Distribution of the LAT parameter as a function of incident electron 
%          or positron energy.  Each plot represents a 1 GeV bin.  The histograms
%          represent a simulated sample of Bhabha events and the points are 
%          taken from \babar\ data.} \label{shower}
% \end{figure*}

% \begin{figure*}[b]
\begin{figure*}
\centering
\includegraphics[width=150mm, angle=0]{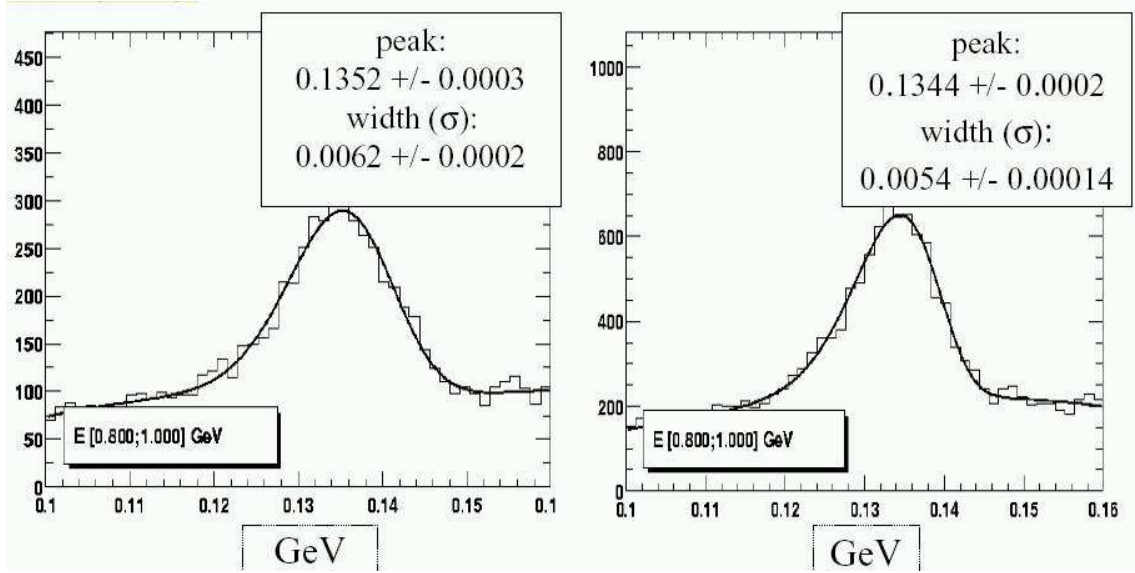}
\caption{$\pi^0$ mass reconstruction for data (left) and simulation (right).
         Histograms represent the reconstructed samples and the curves
         are fits to the samples.  The total pion energy is between 0.8 and
         1.0 GeV.  The fitted mass and width values are stable over
         the range 0.3 - 2.1 GeV.} \label{pi0}
\end{figure*}

\begin{figure*}
\centering
\includegraphics[width=145mm, angle=-90]{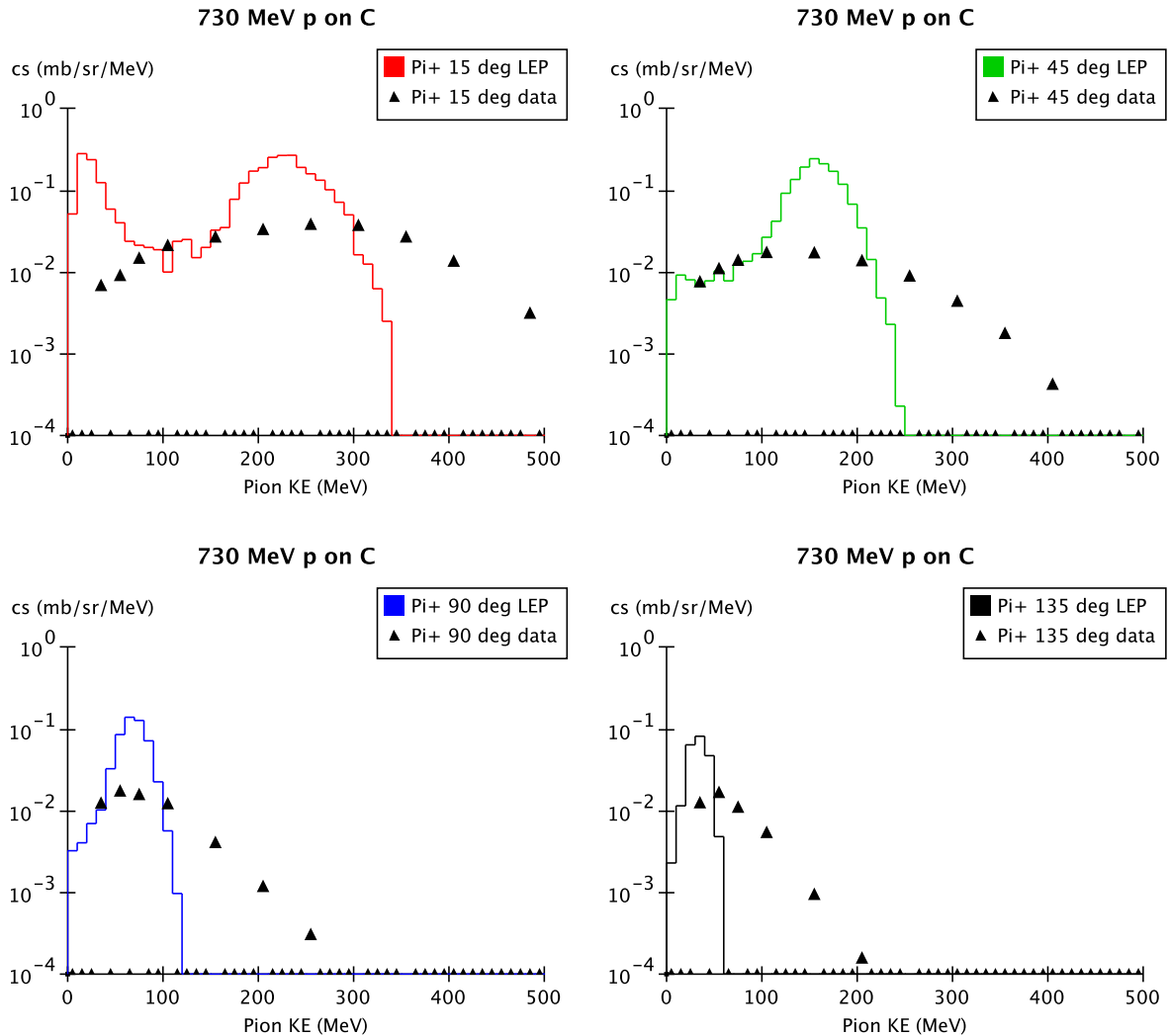}
\caption{Cross section for $\pi^+$ production from 730 MeV protons on carbon
         at $15^{\circ}$, $45^{\circ}$, $90^{\circ}$ and $135^{\circ}$.  
         Histogram: prediction of the Low Energy Parameterized (LEP) model 
         in Geant4.  Data points are from Ref.~\cite{Cochran}.} \label{piLEP}
\end{figure*}

\begin{figure*}
\centering
\includegraphics[width=145mm, angle=-90]{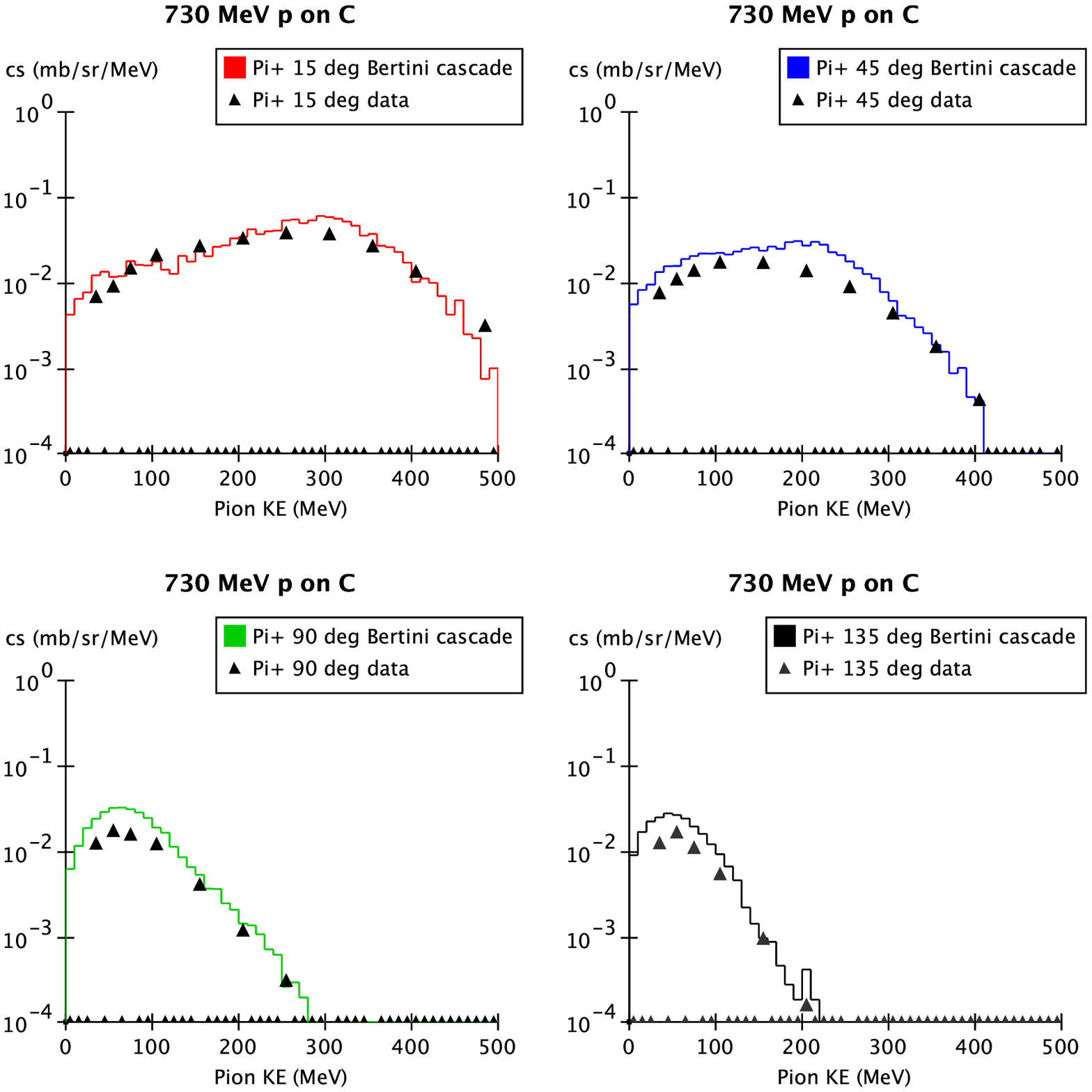}
\caption{Cross section for $\pi^+$ production from 730 MeV protons on carbon
         at $15^{\circ}$, $45^{\circ}$, $90^{\circ}$ and $135^{\circ}$.  
         Histogram: prediction of the Bertini cascade model in Geant4.  Data 
         points are from Ref.~\cite{Cochran}.} \label{piBer}
\end{figure*}

\subsection{Electromagnetic Process Validation}

The \babar\ simulation uses the following Geant4 electromagnetic processes
\begin{itemize}
\item for photons: photo-electric effect, Compton scattering and $e^+ e^-$ 
      pair production;
\item for electrons and positrons: electron ionization, electron 
      bremsstrahlung and $e^+$ annihilation;
\item for muons: $\mu$ ionization, $\mu$ bremsstrahlung and $e^+ e^-$ pair 
      production by muons;
\item for hadrons: hadron ionization;
\item for all charged particles: multiple scattering.
\end{itemize}

One of the most basic validation tests is to reproduce the expected energy 
loss due to ionization processes.  In this case, minimum ionizing electrons 
and positrons from radiative Bhabha scattering were examined in the 
He-Isobutane gas of the drift chamber.  Simulated events are compared 
to data from a recent run in Fig.~\ref{dedx}, where it is seen that both the 
shape and normalization of the $dE/dx$ distributions are in agreement to 
within 15\%.  This indicates that fluctuations in the energy loss are 
reasonably well-reproduced.  More importantly, the mean values of $dE/dx$ 
agree to within 1\%.

It is also important to study electromagnetic showers in the CsI calorimeter.
Measuring shower shapes provides a cumulative test of the photon and electron
processes listed above.  Radiative Bhabha scattering was used to examine
the distribution of azimuthal and lateral shower parameters as a function of
electron/positron energy.  Fig.~\ref{shower} shows the lateral distribution 
parameter (LAT).  With the exception of a few energy bins, 
distributions of both of these parameters were found to show good agreement 
between data and simulation for all electron and positron energies up to 
4 GeV.  

\subsection{Reconstruction}
The comparison of reconstructed events from data and simulation provides a 
stringent validation test.  In \babar\, an accurate reconstruction of the 
$\pi^0$ mass and width provides a test of many detector performance features, 
including tracking, energy scale, shower development, shower containment and
detector response.

In $B^0 \overline{B^0}$ events, the decay $K_s \rightarrow \pi^0 \pi^0$
provides $\pi^0$s with energies between 0.3 and 2.1 GeV.  Fig.~\ref{pi0}
shows the reconstructed mass for $\pi^0$s with total energies between 0.8 and 
1.0 GeV.  For both data and simulated samples the mass is close to the 
expected value, which indicates that tracking, energy scale and shower 
containment are well-understood.  The widths, however, are significantly 
different and point to problems in shower development or detector response.
Since it is believed that the shower development is understood, detector
response should be examined in future validation tests. 

\subsection{Hadronic Process Validation}
The electromagnetic processes of hadron ionization and multiple scattering 
are of primary importance in the propagation of long-lived hadrons through the
detector.  Hadronic interactions are still important, however, because 
scattering from nuclei and the production of hadronic secondaries in the dense
material of the EMC and IFR can affect the energy deposition.

Hadronic interactions in the \babar\ simulation are currently handled by the
so-called ``low energy parameterized'', or LEP, model.  This is a re-engineered
version of the GHEISHA code \cite{GHEISHA}, which has been tuned for 
hadrons with incident energies below 20 GeV.  Even so, it is not particularly 
appropriate at \babar\ hadron energies which are typically below 1 GeV.  This
is demonstrated in Fig.~\ref{piLEP} which shows the pion production cross 
section for 730 MeV protons on a thin carbon target.  Large, qualitative 
differences between the LEP model and data are evident.  Recently, however, 
better models, such as the Bertini cascade \cite{Bertini}, have become 
available.  As shown in Fig.~\ref{piBer}, much better agreement with the data 
is achieved at all angles.  A current limitation of this model is that it
is not valid for kaons.

The hadronic process validation in \babar\ has so far concentrated on the
comparison of results from various models to published data from other 
experiments in which the incident hadron energies are in the \babar\ range.
Validations using \babar\ data are just beginning.  

While \babar\ is certainly not optimized for hadronic physics validation, two 
regions of the detector are useful for tests.  The beam pipe support tube
is a cylinder of carbon fiber and the inner wall of the drift chamber is
a cylinder of beryllium.  Tests are just beginning which use these regions
as thin targets for incident hadrons.

\section{PERFORMANCE}
  The simulation stage of \babar\ event production includes event generation,
tracking and hit-scoring.  A Pentium III 866 MHz PC currently requires 
5.0 seconds on average to simulate a $B^0 \overline{B^0}$ event.  Average 
execution times for other \babar\ event types are given in Table~\ref{table1}.

\begin{table}
\begin{center}
\caption{Execution Time for \babar\ Events}
\begin{tabular}{|l|c|}
\hline \textbf{Event type} & \textbf{CPU time (sec)} \\
\hline $B0 \overline{B0}$ & 5.0 \\
\hline bhabha & 7.0 \\
\hline dimuon & 0.6 \\
\hline
\end{tabular}
\label{table1}
\end{center}
\end{table}

Up to now, relatively little effort has been devoted to optimizing performance
because validation was the main concern.  However, some increase in speed
is expected with

\begin{itemize}
\item improvements in \babar\ geometry models, 
\item improvements in the Geant4 electromagnetic processes, and 
\item fine tuning of the secondary particle production cuts.
\end{itemize}
On the other hand, the improved hadronic models are much more CPU-intensive
than the currently used LEP model, and will cause some slowdown.  

  After the production of 20 million validation events, many of the bugs in
both the \babar\ simulation code and in Geant4 have been shaken out.  To date,
\babar\ simulation production has generated more than 1.5 billion events of 
all types at roughly 20 production sites in North America and Europe, with a 
failure rate of less than one event per million in the simulation stage.  

\section{CONCLUSION}

  \babar\ is the first large experiment to develop and use a Geant4-based 
simulation.  A detailed detector model has been developed using the Geant4
geometry, and both electromagnetic and hadronic interactions are implemented
with Geant4 physics processes.  The simulation was designed to run within
the \babar\ Framework application.  

Comparisons of simulated event samples and \babar\ data indicate that the 
electromagnetic processes perform largely as expected in the few GeV energy 
range.  Validation tests of the LEP hadronic model show that more detailed 
models will be required before good agreement with data can be achieved.

The \babar\ simulation has proven to be robust, with 1.5 billion events 
generated so far at a low failure rate.  Future efforts will concentrate
on improving the overall speed of the simulation. \\

\section{ACKNOWLEDGMENTS}

This work was supported by Department of Energy contract DE-AC03-76SF00515. \\

\end{document}